\documentstyle[aps,preprint]{revtex}
\begin{document}
\tighten
\title{\bf Critical Fluctuations in 
Topologically Massive  
Superconductors}

\author{A. P. C. Malbouisson
\footnote[1]{e-mail: adolfo@lafexsu1.lafex.cbpf.br}, 
F. S. Nogueira\footnote[2]{e-mail: flaviono@cbpfsu1.cat.cbpf.br}
 and N. F. Svaiter\footnote[3]{e-mail: nfuxsvai@lca1.drp.cbpf.br}}

\address{Centro Brasileiro de Pesquisas Fisicas - CBPF,
Rua Dr. Xavier Sigaud 150, Rio de Janeiro, RJ 22290-180, 
BRAZIL}

\date{Received \today}

\maketitle

\begin{abstract}
We consider a topologically massive Ginzburg-Landau model 
of superconductivity. In the context of a mean field calculation, 
we show that there is an increase in the critical temperature 
driven by the topological term. It is shown that this effect 
persists even if we take into account the critical 
fluctuations. 
The renormalization group analysis gives further insight on 
this behavior. The fixed point structure is such that the 
critical exponents tend to their mean field values for 
very large values of the topological mass. In this sense, 
the topological term stabilizes the critical fluctuations of the order 
parameter.
\end{abstract}\pacs{74.20.De, 11.10.Hi}

The Ginzburg-Landau (GL) theory of superconductivity 
\cite{Ginzburg} describes 
very well the phenomenology of conventional superconductors. 
It is believed that a similar phenomenological theory can be 
applied to the study of the High temperature superconductors 
(HTSC). This expectation relies on  
experimental observation which indicates that the order parameter is 
just the same in both situations 
\cite{IBM}. However, the GL theory neglects the 
fluctuations of the order parameter which
 are very important in the HTSC. 
Consequently, the exponents of the HTSC phase transition differs 
 from the ones given by the GL theory \cite{Lobb}.
Theoretically, the fluctuations can be taken into account 
through the use of 
renormalization group techniques to study the behavior of the 
theory in the neighbourhood of the critical point 
\cite{HLM}. Another possible path to study the effect of 
the fluctuations is to compute further corrections to the free 
energy functional in a systematic way by performing a loop expansion 
\cite{Kleinert}. 

In this note we study a version of the GL free energy functional 
where a topological Chern-Simons term \cite{Deser} is added. 
Topological models are frequently employed in the 
construction of quantum models for HTSC which 
explore the effects of statistical transmutation 
(anyon superconductivity) \cite{Fradkin}. Here 
we investigate the effect of such a topological contribution in a 
macroscopic model which generalizes the GL model. We perform  
mean field calculations very similar to those in the 
 mean field theory proposed 
by Halperin, Lubensky and Ma (HLM) \cite{HLMa}.    
Also, we compute the corrections due to the critical 
fluctuations in the free energy and we obtain the renormaliztion 
group equations.

Our starting point 
is the following free energy functional,
 
\begin{eqnarray}
F[\psi,\vec{A}]&=&{\int}d^{3}x\left[\frac{1}{2}|(\nabla-iq\vec{A})\psi|^{2}
+\frac{r_{0}}{2}|\psi|^{2}+\frac{u_{0}}{4!}|\psi|^{4}\right. \nonumber \\
&+&\left.\frac{1}{2}(\nabla\times\vec{A})^{2}+i\frac{\theta}{2}
\vec{A}\cdot(\nabla\times\vec{A})\right],
\end{eqnarray}
where $r_{0}=a_{0}(T-T_{0})/T_{0}$ and $\theta>0$ is the topological 
mass. The partition function is given by

\begin{equation}
Z={\int}D\vec{A}D\psi^{\dag}D\psi\exp(-F[\psi,\vec{A}]).
\end{equation} 
           Since $F$ is quadratic in the vector field, the integration over 
$\vec{A}$ is Gaussian and can be performed exactly. For 
an uniform $\psi$ this defines the following 
 effective free energy density functional:

\begin{equation}
f_{eff}[\psi]=-\frac{1}{12\pi}[M_{+}^{3}(|\psi|^{2})+
M_{-}^{3}(|\psi|^{2})]+\frac{r}{2}|\psi|^{2}
+\frac{u_{0}}{4!}|\psi|^{4},
\end{equation}
where the calculations were performed in the the Landau gauge and 
the functions $M_{\pm}$ are defined by

\begin{equation}
M_{\pm}^{2}(|\psi|^{2})=q^{2}|\psi|^{2}+\frac{\theta^{2}}{2}\pm
\frac{\theta}{2}\sqrt{\theta^{2}+4q^{2}|\psi|^{2}}.
\end{equation} 
The parameter $r_{0}$ has been renormalized to $r=a(T-T_{c})/T_{c}$. 
When $\theta=0$ Eq.(3) reduces to the effective functional obtained 
by HLM. The critical temperature $T_{c}$ does not depend on 
$\theta$ and is the same as in the HLM paper. 

The inverse of the susceptibility is obtained for temperatures above 
the critical temperature by

\begin{equation}
\chi^{-1}=\frac{\partial^{2}f_{eff}}{\partial|\psi|^{2}}
_{|\psi|=0}.
\end{equation}
The critical temperature is defined by the divergence of the 
susceptibility and we get from Eq.(5) the $\theta$-dependent 
critical temperature

\begin{equation}
\tilde{T}_{c}=T_{c}\left(1+\frac{q^{2}\theta}{2{\pi}a}\right).
\end{equation}
The susceptibility has a critical behavior with critical 
exponent $\gamma=1$ which is the mean field value. 
Minimizing the effective free energy functional with respect to 
$|\psi|^{2}$ and denoting by $\sigma$ the corresponding value of 
$|\psi|$ which minimizes $f_{eff}$ we get the 
expression for the mean field order 
parameter,

\begin{equation}
\sigma=\sqrt{\frac{6\tilde{a}}{u_{0}}}\left(1-\frac{T}{\tilde{T}_{c}}
\right)^{1/2},
\end{equation}
where

\begin{equation}
\tilde{a}=a+\frac{q^{2}\theta}{2\pi}.
\end{equation}  
Therefore the 
mean field critical exponents are the same as in HLM as it should be. 
In particular, the above results imply that $\beta=\nu=1/2$. However,  the critical temperature has 
been increased by a factor of $1+q^{2}\theta/2\pi^{2}a$ with respect 
to the mean field critical temperature of HLM. The mean field 
theory of HLM renormalizes 
negligibly the critical temperature while in the present case this is not 
necessarily true because we may have arbitrary values of $\theta$ which 
may increase considerably the critical temperature of the superconducting 
phase transition.  

Up to now we have taken into account only the fluctuations which 
does not change the mean field behavior of the model.
To proceed let us investigate the effects of the critical fluctuations. 
These are obtained by computing the one loop correction arising from 
the fluctuations of the order parameter. The corrected free energy 
density is given by

\begin{eqnarray}
f_{eff}^{1-loop}&=&\frac{r_{0}}{2}|\psi|^{2}+\frac{u_{0}}{4!}|\psi|^{4}
+\frac{1}{4\pi^{2}}\{\int_{0}^{\Lambda}dpp^{2}[\ln(p^{2}+
M_{+}^{2}(|\psi|^{2})) \\ \nonumber
&+&\ln(p^{2}+M_{-}^{2}(|\psi|^{2}))+\ln(p^{2}+r_{0}+u_{0}|\psi|^{2}/2)
-3\ln(p^{2})]\},
\end{eqnarray}
from which we get the corrected inverse susceptibility
for $T>T'_{c}$, $T'_{c}$ being the new critical temperature:

\begin{equation}
\chi_{1-loop}^{-1}=\frac{a}{T_{c}}(T-\tilde{T}_{c})+
\frac{u_{0}}{4\pi^{2}}\int_{0}^{\Lambda}dp\frac{p^{2}}{p^{2}+r'}.
\end{equation}
In the above equation we have replaced $r_{0}$ by 
$r'=\alpha(T-T'_{c})=\chi_{1-loop}^{-1}$ 
since the error involved is beyond one loop order.  
The critical temperature is

\begin{equation}
T'_{c}(\theta)=\tilde{T}_{c}(\theta)-\frac{u_{0}T_{c}
\Lambda}{4\pi^{2}a}.
\end{equation}
Thus, the critical fluctuations have the effect of lowering the 
critical temperature.
However, $T'_{c}(0)<T'_{c}(\theta)$ and therefore the topological term 
increases the degree of order with respect to the standard GL model. 
This effect is better elucidated if one uses Eq.(6) to rewrite (11) 
as

\begin{equation}
T'_{c}(\theta)=T_{c}+\frac{\Lambda{T_{c}}}{2\pi{a}}\left(
q^{2}\overline{\theta}-\frac{u_{0}}{2}\right),
\end{equation}
where we have chosen to measure $\theta$ in units of $\Lambda$, 
that is, $\theta=\Lambda\overline{\theta}$. From Eq. (12) we see 
that the effect of the 
critical fluctuations can be supressed by the topological 
mass. In fact, if $u_{0}=2q^{2}\overline{\theta}$ we have 
$T'_{c}=T_{c}$. Thus, the topological term acts as a factor 
which compensates the disorder introduced by the critical 
fluctuations. By attributing typical BCS values to both 
$q^{2}$ and $u_{0}$ in Eq.(12) we get that
$\overline{\theta}=u_{0}/2q^{2}{\sim}10^{-2}$ in order to 
suppress the effects of the fluctuations of the order parameter 
on the critical temperature. Since 
for a typical BCS situation $u_{0}/2q^{2}{\sim}10^{-2}$ we may choose 
$\overline{\theta}$ such that $\overline{\theta}>>
u_{0}/2q^{2}$ and $T'_{c}\approx\tilde{T}_{c}$. Therefore, 
in a topologically massive GL model the critical temperature can 
be considerably enhanced even when the critical fluctuations are 
taken into account. The fluctuations arising from the vector 
potential $\vec{A}$ dominates over the fluctuations of the order 
parameter.

The critical behavior is better analysed through renormalization 
group (RG) techniques. The case with $\theta=0$ was already analysed 
by many authors \cite{HLM}. The RG study in the ultraviolet limit 
was carried over in the case of Chern-Simons 
scalar QED without a self-coupling 
of the scalar field and show a trivial behavior of the Chern-Simons 
coupling, at least in the context of perturbation theory \cite{Semenoff}.
We are interested in the infrared behavior
 and, therefore, the ultraviolet 
cutoff is kept fixed. We shall work with Wilson's version of the RG in 
its perturbative form \cite{Wilson}. Although the presence of the 
cutoff spoils gauge invariance, it can be shown that as soon as the 
cutoff is removed gauge invariance is recovered \cite{Ellwanger}. In 
fact, many studies using Wilson's RG are being 
performed in gauge theories \cite{Bonini}. Up to one loop order it is 
possible to perform an $\epsilon$-expansion because the antisymmetric 
part of the vector field propagator does not contribute at this order. 
Alternatively, if we want higher orders we can use dimensional reduction 
techniques to obtain the desired power of $\epsilon$.

The RG equations are obtained by integrating out the fast modes and 
performing an appropriate 
rescaling of the momentum variables and fields. The 
flow equations are better expressed in terms of dimensionless 
parameters defined through $r=\Lambda^{2}\overline{r}$, 
$u=S_{d}^{-1}\Lambda\overline{u}$, 
$f=q^{2}=S_{d}^{-1}\Lambda\overline{f}$ and 
$\theta=\Lambda\overline{\theta}$ where 
$S_{d}=2^{1-d}\pi^{-d/2}/\Gamma(d/2)$. 
 The result is given by

\begin{eqnarray}
\frac{d\overline{r}}{dt}&=&(2-\eta)\overline{r}+
\frac{2\overline{u}}{(1+\overline{r})}+\frac{
3\overline{f}}{(1+\overline{\theta}^2)} \\
\frac{d\overline{u}}{dt}&=&(\epsilon-2\eta)\overline{u}-
\frac{5\overline{u}^{2}}{3(1+\overline{r})^{2}}-
\frac{18\overline{f}^{2}}{(1+\overline{\theta}^{2})^{2}}
 \\
\frac{d\overline{f}}{dt}&=&(\epsilon-\eta_{A})\overline{f}
\end{eqnarray}
where the anomalous dimensions for the scalar and  
gauge fields are given respectively by

\begin{equation}
\eta=-3\frac{\overline{f}}
{(1+\overline{r})(1+\overline{\theta}^{2})}
\end{equation}
and
\begin{equation}
\eta_{A}=\frac{\overline{f}}{3(1+\overline{r})^{2}}.
\end{equation}
There is no flow for the parameter $\overline{\theta}$ as expected 
perturbatively. 
From the above flow equations it is readily seen that all critical 
exponents will depend on $\overline{\theta}$. Therefore, 
the parameter $\overline{\theta}$ drive the 
system into different universality classes. 

The fixed point structure is well known for $\overline{\theta}=0$. 
It is found that the superconducting fixed point 
$(\overline{r}^{*},\overline{u}^{*},\overline{f}^{*})$ has a 
complex value for $\overline{u}^{*}$ if the number of components of 
the order parameter is less than 365.9 \cite{CLN}. This behavior 
is usually interpreted as indicating a first order phase transition
driven by fluctuations. 

However, this behavior is changed for $\overline{\theta}>0$. In this 
situation, real superconducting fixed points are found. Typically,  
we find two physical superconducting fixed points, one with two 
attractive infrared directions and one infrared repulsive while the 
other one has two infrared repulsive directions and one attractive. 
For instance, for $\overline{\theta}=3$ and $\overline{\theta}=10$ 
we find the following fixed points with two infrared attractive 
directions $(-0.47,0.24,0.85)$ and $(-0.4,0.24,1.08)$, respectively. 
This type of behavior confirms the already mentioned fact that 
the topological mass damps the critical fluctuations of the 
order parameter. Indeed, for very large values of $\overline{\theta}$, 
the critical exponents tend to their mean field values. For example, 
for $\overline{\theta}=100$ we find for the 
exponent of the order parameter correlation function, 
the value 
$\eta=-0.0006$. We conclude that the mean field exponents becomes 
exact for very large values of the topological mass.

To summarize, we studied a topologically massive version of 
the GL model. The mean field theory gives a critical temperature 
larger than the usual one by a factor proportional to the 
topological mass. This gives an enhancement of the critical 
temperature driven by topological effects. Even when the 
fluctuations of the order parameter are taken into account 
this enhancement persists and we conclude that the topological 
term has the effect of increase the order of the system. 
In this 
sense the topological term may be thought as an inductor of 
higher critical temperatures in unconventional superconductors. 
The fixed point structure exhibits superconducting fixed points 
with  real values for $\overline{u}^{*}$ if $\overline{\theta}>0$. 
The critical exponents have a $\overline{\theta}$ dependence such 
that they tend to the mean field values for $\overline{\theta}$ 
very large. This means that the Chern-Simons term has the effect of 
stabilize the critical fluctuations.

It is an interesting question if this topological macroscopic theory 
could have some relevance to the HTSC case. In order to discuss 
this case it is 
necessary to improve considerably the proposed model. The HTSC are 
highly anysotropic and we are dealing here with an isotropic 
situation. Also, the coherence length is very small as compared to 
the coherence length of conventional superconductors and the effects 
of the critical fluctuations are very important. This means that a 
more careful study of the RG equations is necessary with an
explicit evaluation of the critical exponents. However, it is 
quite remarkable that the critical region for these materials are 
large as compared to conventional superconductors due to the 
smallness of the correlation lenght. Therefore, Ginzburg-Landau 
like models are suitable for a macroscopic description of 
the superconducting phase transition in HTSC.  

We hope that this note could estimulate further investigation 
on the subject, both from the theoretical and experimental point 
of view.

One of us (F.S.N.) would like to thank Dr. T. M. Bambino for 
useful discussions. This work was supported by the brazilian 
agency CNPq.


\begin{references}
\bibitem{comment1} The exact expression of the superconducting 
fixed points as a function of $\overline{\theta}$ is too 
lenghty to be reproduced here. It can be obtained through 
computer algebra tools.
\end{references}
\end{document}